\documentclass{osa-article}

\journal{oe}

\articletype{Research Article}

\usepackage{lineno}

\begin{document}

\title{Highly-enhanced active beam-wander-correction for free-space quantum communications}

\author{Dohoon Lim,\authormark{1} Dongkyu Kim,\authormark{1} Kyungdeuk Park,\authormark{1} Dong-Gil Im,\authormark{1} and Yong Sup Ihn\authormark{1,*}}

\address{\authormark{1} Emerging Science and Technology Directorate, Agency for Defense Development, Daejeon 34186, South Korea}

\email{\authormark{*}yong0862@add.re.kr} 

 

\begin{abstract*}
In practical applications to free-space quantum communications, the utilization of active beam coupling and stabilization techniques offers notable advantages, particularly when dealing with limited detecting areas or coupling into single-mode fibers(SMFs) to mitigate background noise.
In this work, we introduce highly-enhanced active beam-wander-correction technique, specifically tailored to efficiently couple and stabilize beams into SMFs, particularly in scenarios where initial optical alignment with the SMF is misaligned. 
To achieve this objective, we implement a SMF auto-coupling algorithm and a decoupled stabilization method, effectively and reliably correcting beam wander caused by atmospheric turbulence effects. 
The performance of the proposed technique is thoroughly validated through quantitative measurements of the temporal variation in coupling efficiency(coincidence counts) of a laser beam(entangled photons). 
The results show significant improvements in both mean values and standard deviations of the coupling efficiency, even in the presence of 2.6 km atmospheric turbulence effects.
When utilizing a laser source, the coupling efficiency demonstrates a remarkable mean value increase of over 50 $\%$, accompanied by a substantial 4.4-fold improvement in the standard deviation.
For the entangled photon source, a fine mean value increase of 14 $\%$ and an approximate 2-fold improvement in the standard deviation are observed.
Furthermore,the proposed technique successfully restores the fidelity of the polarization-entangled state, which has been compromised by atmospheric effects in the free-space channel, to a level close to the fidelity measured directly from the source.
Our work will be helpful in designing spatial light-fiber coupling system not only for free-space quantum communications but also for high-speed laser communications.
\end{abstract*}


\section{Introduction}

The field of quantum key distribution(QKD) has witnessed remarkable advancements in achieving long-range transmissions since its initial demonstration in 1992 \cite{JCrypt92Bennett}.
Fiber-based links have successfully demonstrated transmission distances of up to 1000 km in twin-field QKD \cite{PRL23Liu} and up to 442 km using measurement-device-independent QKD \cite{PRA23Liu}. 
On the other hand, terrestrial free-space links have shown impressive capabilities up to 144 km \cite{Nature07Ursin,Nature12Yin}.
Recent advancements in free-space links have been particularly noteworthy, making pioneering breakthroughs in the field. 
Notably, QKD through satellite-ground free-space links has become a well-established technology \cite{Science17Yin, Nature17Liao, RMP22Lu}. 
Furthermore, QKD over air-to-ground links utilizing unmanned air vehicles (UAVs) and aircrafts has also achieved notable progress\cite{NPhoton13Nauerth, NSR20Liu, IEEE21Alshaer}. 
In addition to these accomplishments, urban QKD systems designed for operation in daylight conditions have seen substantial development \cite{AO13GarciaMartinez,SR18Ko}.
Despite these impressive achievements, the practical implementation of reliable free-space quantum communication still faces substantial technical challenges related to background noise and atmospheric effects.

One effective technique to mitigate the impact of background noise in a quantum communication system is to reduce the field of view of the receiver's detectors.
Especially when dealing with small detecting areas or coupling into a single-mode fiber(SMF), efficient and stable light coupling into the optical fiber is crucial to ensure communication efficiency and the success rate of the receiving system.
The process of light coupling into an optical fiber is influenced by beam wander, platform vibration, and misalignment between the fiber and focus spot.
Here, atmospheric temperature variations, humidity gradients, and fast-varying atmospheric turbulence can cause beam wander, beam flickering, and wavefront distortion, leading to random fluctuations of the received beam and the apparent azimuth of the receiver over time scales ranging from milliseconds to minutes, resulting in increased system losses \cite{Nature07Ursin,PRL07Schmitt-Manderbach,OE14Carrasco-Casado,MOTL16Carrasco-Casado,CAP22Kim}.

To address these challenges, active beam-wander-correction techniques plays a crucial role in minimizing losses and achieving efficient coupling and stabilization of beams into an optical fiber coupling system, although the utilization of adaptive optics techniques compensating wavefront error is also needed for further decreasing optical losses.
Various excellent methods have been proposed so far for beam stabilization and wavefront correction \cite{SPIE02Weyrauch,AO15Chen,SPIE17Gruneisen,OE20Yang,JLT21Mai,OC23Mai}.
However, there has been no research demonstrating the performance of beam-wander-correction techniques in a FSO setup based on entangled photon-pairs, where the coupling performance of the photon-pairs can be directly monitored.
In this study, we focus on the beam-wander-correction technique of minimizing beam path fluctuation caused by atmospheric turbulence through a 60-m indoor experiment, primarily aiming for stable beam coupling into a SMF.
Such investigation is crucial for exploring novel and efficient approaches to tackle the challenges posed by atmospheric effects in free-space quantum communication systems.

Active beam-wander-corrections involve real-time adjustments and manipulation of the beam path to compensate for environmental factors or imperfections in the optical setup.
By actively monitoring and dynamically adjusting the beam path, it becomes possible to mitigate losses caused by factors such as misalignment, atmospheric turbulence, or thermal effects.
In previous free-space quantum communication setups, the predominant approach for mitigating beam wander and platform jitter effects at the receiver has relied on the utilization of a single fast-steering mirror(FSM).
However, the limitation of employing a sole FSM lies in its capacity to stabilize the beam path exclusively along a single plane of the optical axis \cite{NP13Wang,NP17Liao,OE18Gong,IEEE18Fernandez}.
Particularly, when the transmitting or receiving module moves or when the direction of the incoming beam deviates at the receiver, the coupling efficiency may significantly decrease.
Therefore, to achieve beam wander and jitter correction for all beam paths corresponding to the entire optical axis, the implementation of multiple FSMs is required, along with the development of algorithms specifically designed for this purpose.

In this work, we present highly-enhanced active beam-wander-correction technique using two FSMs and two position-sensitive-detectors(PSDs) configuration, alongside an SMF auto-coupling algorithm and a decoupled stabilization method.
This approach is particularly applicable in scenarios where the initial optical alignment deviates from the SMF.
To effectively simulate atmospheric effects in long-range outdoor environments, we collected beam wander data at a distance of 2.6 km and constructed a vibration simulator by incorporating an additional FSM at the transmitter.
Here, due to spatial constraints within our research facility, the simulated distance is limited to 2.6 km.

By subjecting the system to these simulated atmospheric effects, we rigorously tested the performance of the proposed method, closely monitoring the SMF-coupling efficiency (coincidence counts) for both a laser source and an entangled photon source.
Notably, our experimental results demonstrate very stable coupling efficiency, and we observed a remarkable reduction of its fluctuation by a factor of 4.4 for the laser source and a factor of 2 for the entangled photon source, as compared to the case without correction.
The results provide valuable insights into the design of SMF-coupled optical links for both free-space quantum communications and high-speed laser communications applications.
The novel combination of double FSM and PSD configuration, SMF auto-coupling algorithm, and decoupled stabilization method has the potential to significantly enhance the reliability and performance of free-space quantum communication systems and advancements in high-speed laser communication technologies.

\section{Auto-coupling algorithm and decoupled stabilization method}

\begin{figure}[t]
\centering
\includegraphics[width=0.7\linewidth]{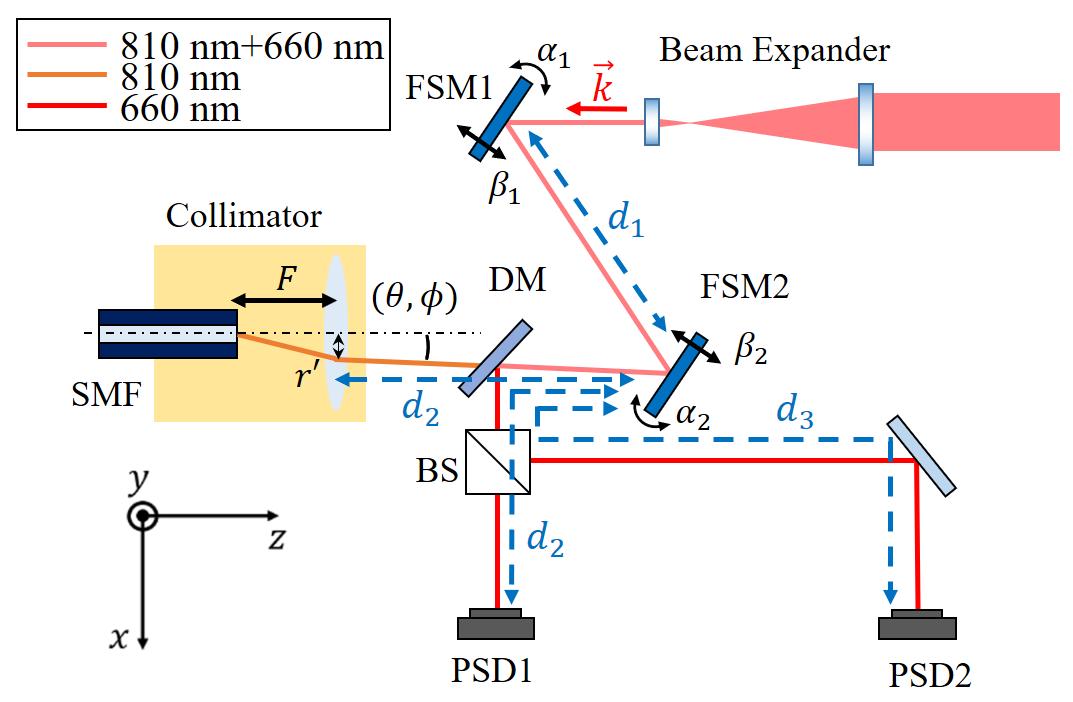}
\caption{The schematic diagram of the SMF auto-coupling system, employing double FSM configuration. SMF, single-mode fiber; FSM, fast-steering mirror; DM, dichroic mirror; PSD, position sensitive detector.}
\label{fig1}
\end{figure}

Figure \ref{fig1} illustrates the schematic of the SMF auto-coupling system employing two FSMs.
The system effectively controls the angles of the two FSMs, which are determined by their horizontal components ($\alpha_1$,$\alpha_2$) and vertical components ($\beta_1$,$\beta_2$).
To achieve precise beam alignment, a collimated beam is directed towards FSM1 to scan the angle of the optical beam path and subsequently reflected to FSM2 to scan the radial position of the beam spot.
The beam is then focused by a collimator with a focal length $F$ and finally coupled into the SMF.
The SMF coupling efficiency $\eta$ can be defined as the ratio of the power $P_c$ coupled into the SMF to the input power $P_i$ \cite{OFT21Cao,SPIE98Ruilier,book04Buck}:
\begin{equation}
\eta= \frac{P_c}{P_i}=\frac{\left\vert \int\int E^{*}(x,y)\cdot F_{G}(x,y)dxdy\right\vert^2}{\int\int\left\vert E(x,y)\right\vert^{2}dxdy},
\label{eq1}
\end{equation}
where $E(x,y)$ and $F_{G}(x,y)$ refer to the incident electromagnetic field and the Gaussian field distribution of the SMF, respectively.
If an optical beam path has misaligned with the normal axis of the SMF, the beam propagating direction $\vec{k}$ undergoes a radial position shift $r'=\sqrt{x'^{2}+y'^{2}}$ and an angular offset $\theta$, defined as the angle between $\vec{k}$ and $z$-axis, at the collimator lens.
In this case, the incident Gaussian field with the waist radius of $w_s$ can be given by 
\begin{align} \begin{split}
E(x,y)&= \sqrt{\frac{2}{\pi w^{2}_s}}\; \text{exp}\left[-\frac{(x-x')^{2}+(y-y')^{2}}{w^{2}_s}\right]\\
&\hspace{5mm} \times \text{exp}[ik\theta (x\text{cos}\phi+y\text{sin}\phi)],
\label{eq2} \end{split}
\end{align}
where $\phi$ is the azimuthal angle of $\vec{k}_{\bot}$ which is the projection of $\vec{k}$ into the $xy$-plane. 
In addition to the incident Gaussian field, the fiber's Gaussian distribution with a mode-field radius of $w_0$ can be expressed in
\begin{equation}
F_{G}(x,y)= \sqrt{\frac{2}{\pi w^{2}_m}}\; \text{exp}\left[-\frac{x^{2}+y^{2}}{w^{2}_m}\right],
\label{eq3}
\end{equation}
where $w_{m}=\frac{\lambda F}{\pi w_{0}}$ is the effective mode-field radius, $F$ is the focal length of the collimator, and $\lambda$ is the wavelength of the incident Gaussian beam.
If we assume that the aperture radius of the collimator is much larger than the Gaussian beam size and the angular offset $\theta$ is very small, the coupling efficiency $\eta$ of the SMF can be calculated as
\begin{equation}
\eta = \biggl[\frac{2 w_m w_s}{w_m^2+w_s^2}\; \exp\biggl(-\frac{4r'^2+k^2{\gamma}^2 w_m^2 w_s^2}{4(w_m^2+w_s^2)}\biggr)\biggr]^2
\label{eq4}
\end{equation}
where $\gamma = \arccos{(\cos{\theta}\cos{\phi})}$ and $k=2\pi/\lambda$ is the wave vector.
Here, the radial position shift, $r'=\sqrt{r^{2}_{\text{H}}+r^{2}_{\text{V}}}$, consists of the horizontal component, $r_{\text{H}}=2d_1\alpha_1+2d_2(\alpha_1-\alpha_2)$, and the vertical component, $r_{\text{V}}=2d_1\beta_1+2d_2(\beta_1-\beta_2)$, where $d_1$ refers to the distances between FSM1 and FSM2, and $d_2$ denotes the distance between FSM2 and collimator.
The SMF auto-coupling system operates on an angle scan, followed by the position scan, employing two FSMs.
During the angle scan, FSM1 performs random scans of horizontal and vertical angles, namely $\alpha_1$ and $\beta_1$, while keeping $\alpha_2$ and $\beta_2$ fixed.
Subsequently, the simultaneous scanning of $\alpha_1$ and $\alpha_2$ ($\beta_1$ and $\beta_2$) takes place, with the constraint that the difference $\alpha= \alpha_1-\alpha_2$ ($\beta= \beta_1-\beta_2$) remains constant, thus facilitating the position search.

\begin{figure}[t]
\centering
\includegraphics[width=0.7\linewidth]{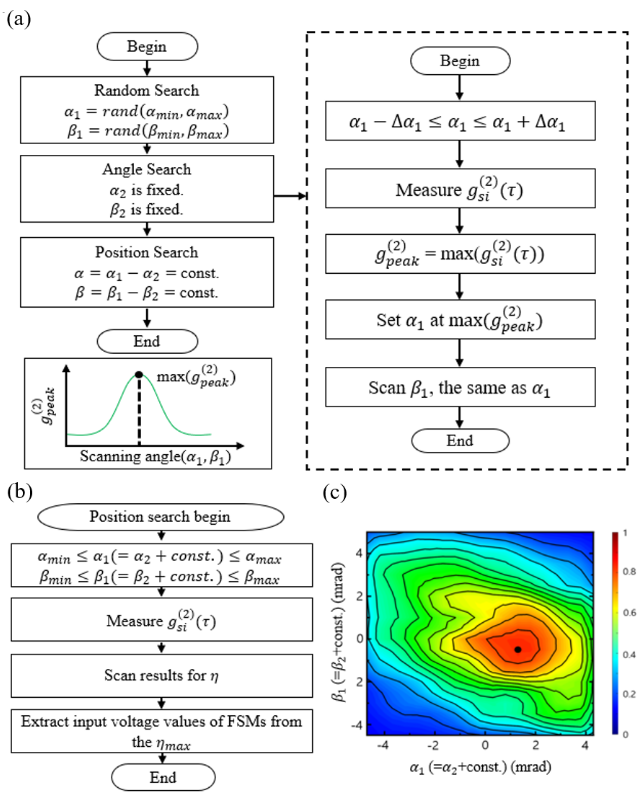}
\caption{Auto-coupling algorithm. (a) Flowchart of the angle search algorithm. The inset shows $g^{(2)}_{peak}$ values as a function of the scanning angle $\alpha_1$ ($\beta_1$) in the flowchart. Here, $\Delta\alpha_1$($\Delta\beta_1$)=1.5 mrad. (b) Flowchart of the position search algorithm. (c) Position scan result for the fiber-coupling efficiency $\eta$ as a function of $\alpha_1$ and $\beta_1$. The solid circle indicates the point of the maximum coupling efficiency $\eta_{max}$.}
\label{fig2}
\end{figure}

At the receiver end, the optical setup comprises two FSMs (S-330.4SL, Physik Instrumente), exhibiting an angular resolution of 0.25 $\mu$rad.
A fiber collimator (60FC-4-M8-10, Sch$\ddot{\text{a}}$fter $+$ Kirchhoff) with a focal length $F$ of 8.1 mm and an SMF (780HP, Thorlabs) with a mode-field radius ($w_{0}$) of 2.5 $\mu$m are also incorporated. 
Notably, system parameters such as the wavelength $\lambda$ are set to 810 nm, while the incident Gaussian field possesses a waist radius ($w_s$) of 0.6 mm.
The distances $d_1$, $d_2$, and $d_3$ are specified as 9 cm, 28 cm, and 76 cm, respectively.
To achieve precise angle variations, a relatively large path difference between PSD1 and PSD2 (C10443-03, Hamamatsu) has been intentionally established.

The auto-coupling algorithm encompasses three primary stages as visually depicted in Fig. \ref{fig2}(a) and (b).
Initially, the random search phase involves the random scanning of $\alpha_{1}$ and $\beta_1$ to ascertain the position where the second-order correlation function $g^{(2)}_{si}(\tau)$ of coupled signal photons in the receiver and directly measured idler photons from the source (see Fig. \ref{fig4}), or the laser beam power, is measured from the initial alignment position, which lies outside the confines of the SMF.
In our system, the angular range of $\alpha_{1}$ and $\beta_1$ spans from $-$5 to 5 mrad.
To achieve the randomness, the input voltage for FSM1 is varied within the range of 0 to 10 V, with the randomness being generated through the LabView graphical programming environment.

Subsequently, the angle searching process proceeds in sequential manner, wherein $\alpha_{1}$ and $\beta_1$ are scanned to identify the optimum values within a narrower range compared to the random search phase.
During actual measurements, the scan ranges of $\alpha_{1}$ and $\beta_1$ are reduced from 10 to 3 mrad around the position where $g^{(2)}_{si}(\tau)$ can be measured through random search process starting from a completely misaligned initial position, and at that time, the step size is set to 100 $\mu$rad.
At each step of the scanning process, $g^{(2)}_{si}(\tau)$ and $g^{(2)}_{peak}$, which is the peak value of $g^{(2)}_{si}(\tau)$, are measured, and the maximum $g^{(2)}_{peak}$ value is continually updated based on the current measurements. 
After the angle scanning process, $\alpha_1$ and $\beta_1$ values are set to the point with the maximum $g^{(2)}_{peak}$, and the angle search process is ended.

Figure \ref{fig2}(b) shows the flowchart of the position search algorithm.
In this process, the optimum positions are determined by the simultaneous scan of $\alpha_1$ and $\beta_1$ in steps of 0.5 mrad across the full range ($-$5 to 5 mrad), while the incident angles $\alpha$ $(=\alpha_1-\alpha_2)$ and $\beta$ $(=\beta_1-\beta_2)$ have fixed values. 
Consequently, while the difference in angles between the two FSMs remains constant, the beam's entry angle into the collimator lens remains fixed, while the beam path position at the collimator lens changes correspondingly.
The results of the complete scan are represented in Fig. \ref{fig2}(c) as a contour plot of the coupling efficiency $\eta$, revealing its dependency on $\alpha_1$ and $\beta_1$.
To achieve smoother representation, the measured data underwent smoothing using a Gaussian filter.
Subsequently, based on the scan result, the input voltages were applied to the FSMs, corresponding to the point of the maximum coupling efficiency, $\eta_{max}$.
The proposed algorithm exhibits applicability in both the classical and quantum source schemes.
For laser sources, instead of the second-order correlation function $g^{(2)}_{si}(\tau)$, the optical power $P$ coupled into a SMF was measured. 

\begin{figure}[t]
\centering
\includegraphics[width=0.6\linewidth]{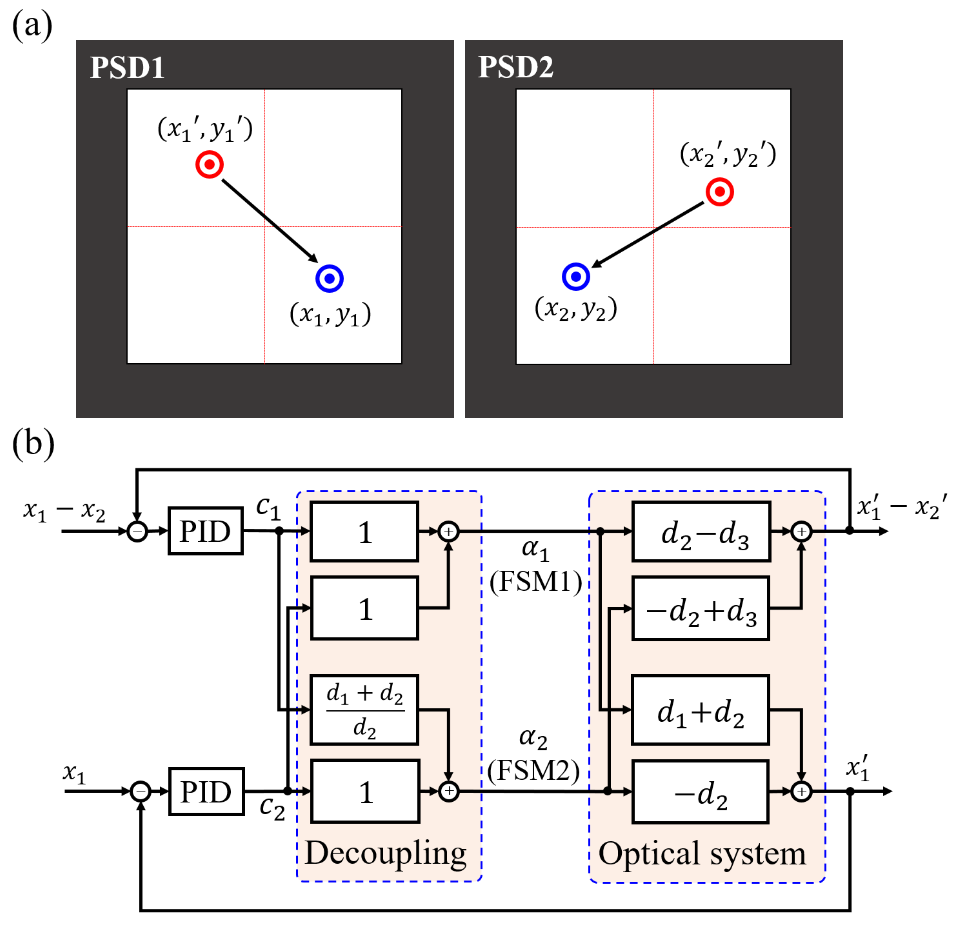}
\caption{ Decoupled beam stabilization method. (a) The detection planes of PSD1 and PSD2, with $(x_1, y_1)$ and $(x_2, y_2)$ representing the beam positions on the respective PSD sensors. These positions serve as reference points acquired from the auto-coupling process, facilitating precise beam alignment and stabilization. (b) The decoupled stabilization system using the PID control algorithm, ensuring independent control of the beam stabilization process. The algorithm is applied to stabilize both $x$- and $y$-components of the laser beam. Notably, the feedback process for the $y$-component remains identical to that of the $x$-component.}
\label{fig3}
\end{figure}

Following the auto-coupling process, a beam stabilization algorithm based on the proportional-integral-differential (PID) control algorithm is established to ensure the stability of the laser beam(or entangled photons) in the laser(or quantum) communication channel.
The stabilization system incorporates two PSDs for accurate beam position measurement.
Leveraging the two FSMs originally employed in the auto-coupling process, the system effectively achieves beam wander stabilization.
To address the rapid fluctuations of beam wander, a common simplified approach involves using a single set of FSM-PSD, where a PID controller is applied to the FSM to maintain the laser beam consistently at the center of the PSD, thereby achieving stabilization. 

However, in our study, we adopt a more elaborate strategy employing two sets of FSM-PSD to enable comprehensive beam-wander-correction for all possible beam paths.
Although previous research has successfully achieved beam stabilization in FSO communication systems by employing two sets of FSM-PSD \cite{OE16Liu, OE21Liang}, the application of the decoupled method proposed in this study, enabling independent control of two FSMs for effective beam stabilization, has not been implemented yet in free-space quantum communication systems. 
In a typical scenario where the PID control method is employed to concurrently manage both FSMs, it exerts an influence on the laser's motion and consequently impacts the measurements obtained from PSD1 and PSD2.
Therefore, this simultaneous control could significantly affect the stability of the feedback process.
Hence, to achieve stable and accurate beam positioning, each PID control loop is designed to be decoupled from one another, guaranteeing that their operations remain independent and do not interfere with each other.

Let's denote the beam positions detected by PSD1 and PSD2 at the point of maximum coupling efficiency point as $(x_1, y_1)$ and $(x_2, y_2)$, respectively, as depicted in Fig. \ref{fig3}(a).
For simplicity, we describe only the $x$-component feedback process, noting that the feedback process for $y$-component is identical.
Figure \ref{fig3}(b) shows the PID control algorithm implemented in the decoupled stabilization method. 
To regulate the incident beam angle on the collimator (Fig. \ref{fig1}), we employ the difference between the position values of PSD1 and PSD2, denoted as $x_1-x_2$.
Additionally, to control the beam position incident on the collimator, we utilize the position value $x_1$ from PSD1, which is situated at the same distance as the collimator.
The feedback process yields output values $x_1'-x_2'$ and $x_1'$, which are subtracted from $x_1-x_2$ and $x_1$, respectively, generating error values that are fed back into the PID controller.
The decoupling process, represented by the matrix $D = \left[\begin{matrix} 1 & 1 \\ \frac{d_1+d_2}{d_2} & 1 \end{matrix}\right]$, is implemented to independently control the PID control values, $c_1$ and $c_2$.
The resulting output values $\alpha_1$ and $\alpha_2$ from the decoupling process serve as inputs to FSM1 and FSM2, respectively, facilitating the adjustment of mirror angles.
These adjusted angles influence the beam path passing through the optical system, leading to output values $x_1'-x_2'$ and $x_1'$.
The optical system can be effectively represented as a transfer function $P= \left[\begin{matrix} d_2-d_3 & -d_2+d_3 \\ d_1+d_2 & -d_2 \end{matrix}\right]$ illustrating the overall control algorithm process.
Upon careful examination of the transfer function, $P\left[\begin{matrix} \alpha_1 \\ \alpha_2 \end{matrix}\right] = \left[\begin{matrix} \frac{(-d_2+d_3)d_1}{d_2} & 0 \\ 0 & d_1 \end{matrix}\right]  \left[\begin{matrix} c_1 \\ c_2 \end{matrix}\right]$, it becomes evident that it solely comprises diagonal terms.
This noteworthy observation indicates that each PID control is adept at independent control of $c_1$ and $c_2$ \cite{AlChEJ70Luyben,JPC06Nordfeldt}.

\section{Experimental setup}
\begin{figure}[b]
\centering
\includegraphics[width=0.6\linewidth]{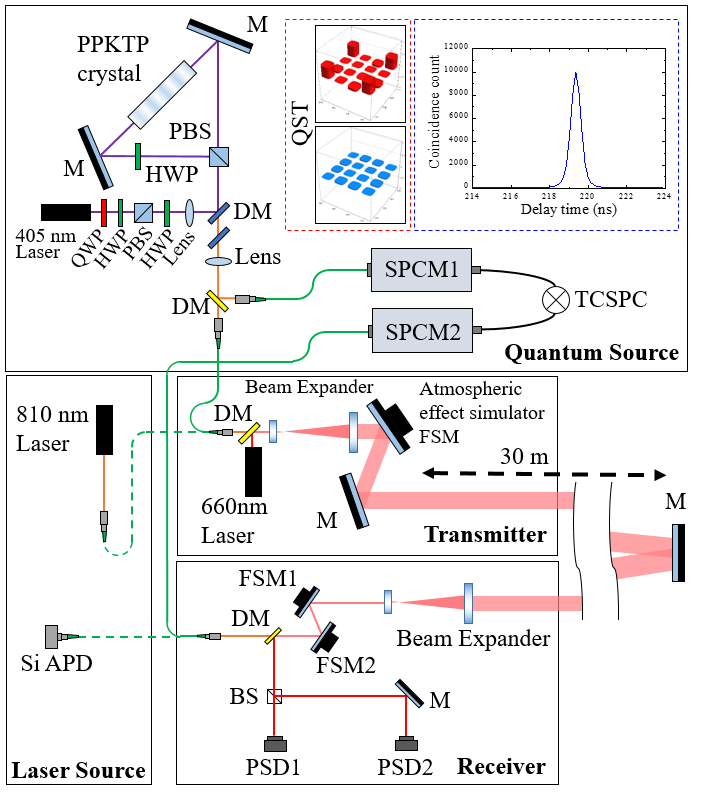}
\caption{Free-space optical set-up with a laser source and a polarization-entangled photon source. The proposed method is evaluated using both quantum and classical sources, incorporating a type-0 spontaneous parametric down conversion (SPDC) photon source and a 810-nm laser beam, respectively. SPCM, single-photon counting module; FSM, fast steering mirror; PSD, position sensitive detector; DM, dichroic mirror; BS, beam splitter; PBS, polarizing beam splitter; HWP, Half wave plate; QWP, Quarter wave plate; M, mirror; QST, quantum state tomography.}
\label{fig4}
\end{figure}

Figure \ref{fig4} shows the experimental setup for the FSO system incorporating beam-wander-corrections based on a double FSM-PSD configuration.
A continuous-wave (CW) laser operating at a wavelength of 810 nm serves as the classical light source.
In contrast, for the quantum source, we generate polarization-entangled photons within a Sagnac interferometer.
Specifically, a single-frequency laser operating at 405 nm pumps a 30 mm-long periodically poled potassium titanyl phosphate (ppKTP) crystal with a 3.425 $\mu$m poling period.
Subsequently, non-degenerated photon-pairs comprising signal ($\sim$780 nm) and idler ($\sim$840 nm) are collinearly generated via a type-0 SPDC process.
The emitted photon-pairs in both clockwise and counter-clockwise directions within the Sagnac interferometer form a polarization-entangled state, represented as $\vert\Phi^{+}\rangle=\frac{1}{\sqrt{2}}(\vert HH \rangle+\vert VV \rangle)$.
Insets in Fig.\ref{fig4} illustrate the density matrix of the two-photon polarization state, exhibiting a fidelity $F$ of 0.97, and the second-order correlation function $g^{(2)}(\tau)$, which determines the rate of coincidence detection between modes $s$ and $i$ at a time delay $\tau$.
Here, the quantum source is operated in a relatively lower pump power regime where fidelity is relatively high, taking into account the multiphoton effects.
The coincidence counts was approximately 9.0 $\times$ 10$^4$ Hz/mW at the transmitter.
A CW laser with a wavelength of 660 nm, referred to as the tracking laser, is employed to monitor and correct beam wander effects in the received beam.

\begin{figure}[t]
\centering
\includegraphics[width=0.7\linewidth]{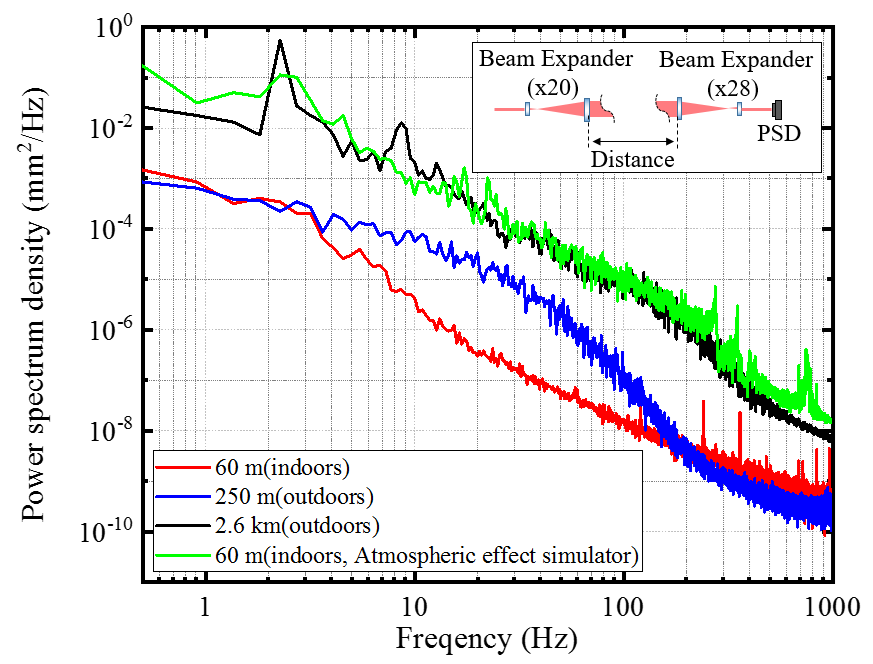}
\caption{Power spectrum density of beam wander due to atmospheric turbulence as a function of frequency for different distances. The measurements were conducted at various distances, namely, 60-m indoors, 250-m outdoors, and 2.6-km outdoors.}
\label{fig5}
\end{figure}

The transmitter units, receiver units, and mirror targets are linked by a free-space channel with a round-trip distance of 60 m.
The transmitter consists of a fiber collimator and an achromatic Galilean beam expander (GBE10-B, Thorlabs) with a fixed magnification of 10 times.
To simulate air turbulence encountered in outdoor environments, two piezo-actuators (P-840, PI) are installed on 2-inch mirror, serving as an atmospheric turbulence simulator. 
By inducing rapid vibrations along the $x-$ and $y-$ axes of the mirror, these actuators introduce significant perturbations to the transmitted light source, effectively emulating external atmospheric turbulence.
The simulated data for the air turbulence is obtained through actual measurements.
The experimental measurement setup, as depicted in the inset of Fig.\ref{fig5}, utilizes a 20$\times$ beam expander at the transmitter and a 28$\times$ beam expander at the receiver to measure the beam wander of the transceiving laser beam using a PSD.
The positions along the $x$ and $y-$axes are measured at a rate of 10 kHz.
Subsequently, the power spectrum density is calculated by averaging 10 measurements with a 0.05 Hz interval to analyze the frequency characteristics of the measured signal, representing the beam wander due to the atmospheric turbulence.

The measurements are conducted at distances of 60-m indoors, 250-m outdoors, and 2.6-km outdoors, respectively.
During the outdoor measurements, the weather conditions are recorded, including a temperature of 26.5 $^\circ$C, wind speed of 2 m/s, humidity of 55 $\%$, and a partially cloudy environment, from 15:00 to 17:00 on 20 June 2023 (Korea Standard Time, KST).
The results demonstrate a pronounced presence of frequency signals in the range of 100 Hz to 300 Hz as the distance increases, with a particularly significant magnitude observed at 2.6 km compared to 250 m.
For the simulation of atmospheric turbulence, the time data of the PSD measurements along the $x$ and $y-$axes at 2.6-km are encoded into the simulator to replicate atmospheric effects experienced at 2.6 km in the 60-m indoor FSO setup.
The results show that the simulated data from the indoor setup and the measured data at 2.6 km exhibit similarity, as corroborated by Fig. \ref{fig5}.
Here, the simulated vibration frequency range emanating from the transmitter spans from 0 to 1000 Hz, with the angular scope of beam wander being marginally smaller than the primary lens of the receiver telescope, encompassing a range from $-$ 0.04 $^{\circ}$ to 0.04 $^{\circ}$.
Given that the ascertained beam wander angle at a distance of 2.6 km amounts to approximately $\pm$0.02 $^{\circ}$, this demonstrates that the 60-m indoor system adequately simulates 2.6 km outdoor conditions.

The receiver part comprises a Galilean telescope (BE05-10-B, Thorlabs), offering variable optical beam expansion with 5 to 10 times magnification, featuring an objective lens of 45 mm diameter.
A collimated beam, 15 mm in diameter, is propagated towards the receiver via a target mirror placed at a distance of 30 m.
To compensate for atmospheric effects, the received beam undergoes a beam stabilization process within the receiver module.
Beam stabilization is achieved using two PSDs and two FSMs.
The transmitter emits either a tracking laser with a wavelength of 660 nm, following the same path as the quantum light source, or a laser source with a wavelength of 810 nm.
Upon reception at the receiver, the 660 nm tracking beam is directed towards the PSD, while the 810 nm beam is directed towards the fiber coupler through the dichroic mirror.
The aforementioned beam stabilization algorithm is implemented using an FPGA (USB-7855, NI), effectively controlling the angles of the FSMs.
The 810 nm beam is ultimately coupled into a SMF, and the coupled optical power within the SMF is evaluated using a fiber-type photodiode.
In the case of employing a SPDC source, the idler photons are directly transmitted to a SPCM, serving as heralding photons for the signal photons.
The signal photons propagate through a 60 m free-space channel, where they are detected by another SPCM, and the coincidence counts are measured using a time-correlation single photon counter (TCSPC).
The peak values of the $g^{(2)}_{si}(\tau)$ functions are used to determine the optimal angle and position of the signal photons, thereby maximizing the coincidence counts.

\section{Results}
\begin{figure}[t]
\centering
\includegraphics[width=0.7\linewidth]{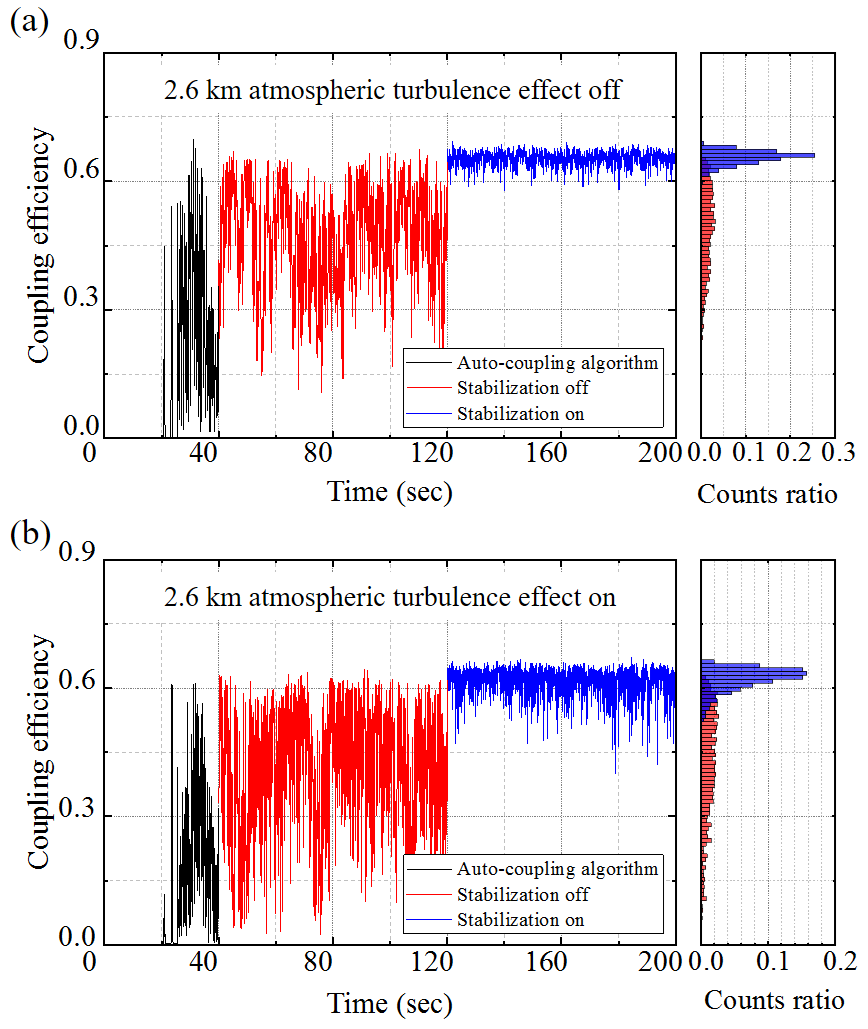}
\caption{Temporal variations of beam coupling efficiencies of an 810-nm laser source with the proposed beam-wander-correction technique applied. (a) The case without atmospheric turbulence effects. (b) The case with 2.6 km atmospheric turbulence effects. The black-colored data denotes the range where auto-coupling algorithm is applied. The red-colored data corresponds to the coupling efficiency after applying the auto-coupling algorithm but before applying the stabilization technique. Finally, the blue-colored data illustrates the coupling efficiency after applying the stabilization technique.}
\label{fig6}
\end{figure}

We first conduct experimental investigations to validate the proposed auto-coupling algorithm and stabilization method within a 60 m FSO setup, employing an 810 nm laser source. 
The free-space channel loss just before the collimator is approximately 16 $\%$.
To measure the coupled optical power, we utilize a fiber-type photodiode (Fig. \ref{fig4}).  
As mentioned earlier, we implement an auto-coupling algorithm to identify the point of maximum coupling power and subsequently proceed with coupling stabilization as shown in Fig. \ref{fig6}.
The data line shown in black represents the process of finding the point of maximum coupling efficiency, starting from an initial state where beam alignment is completely misaligned.
Here, the coupling efficiency denotes the ratio between the laser power measured in front of the receiver's fiber collimator and the coupled laser power into a SMF.
The red-colored data line illustrates significant fluctuations in coupling efficiency due to atmospheric disturbances, particularly near the vicinity of the point where the coupling efficiency is maximized.
Subsequently, employing the decoupled stabilization method yields stable beam coupling into a SMF, as depicted by the blue line.
We apply the decoupled stabilization method to both cases, with and without the influence of atmospheric turbulence effects.
The results indicate that without simulating atmospheric turbulence effects and without applying the stabilization technique, the distribution of coupling efficiency exhibits a mean of 0.471 and a standard deviation of 0.113, as shown in Fig. \ref{fig6}(a).
\begin{figure}[t]
\centering
\includegraphics[width=0.7\linewidth]{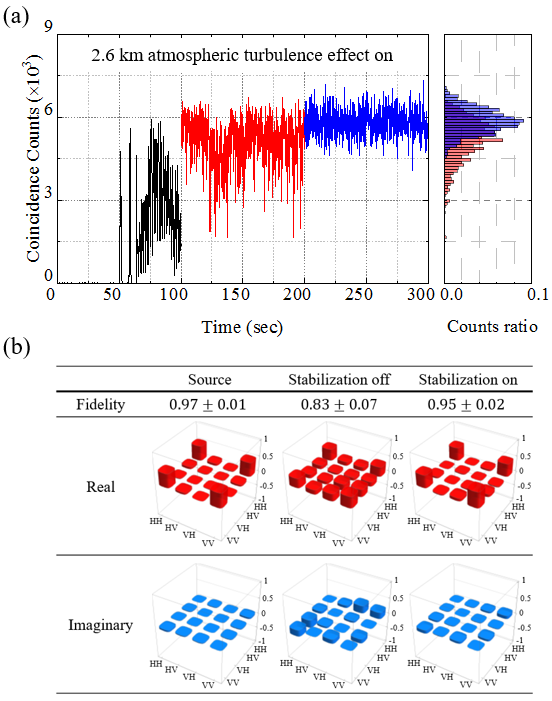}
\caption{(a) Temporal variations of coincidence counts of an entangled photon-pair source with the proposed beam-wander-correction technique applied when subjected to 2.6 km atmospheric turbulence effects. (b) Fidelity variation of polarization-entangled state according to the application of the stabilization method under 2.6 km atmospheric turbulence effects.}
\label{fig7}
\end{figure}
In contrast, when turbulence is not simulated but only stabilization is applied, the mean of the coupling efficiency improves significantly to 0.653, indicating an approximate enhancement of 39 $\%$.
Furthermore, at that point, the corresponding standard deviation is 0.017, representing a considerable improvement of over 6.5 times compared to the case without stabilization.
In Fig. \ref{fig6}(b), which represents the simulation of 2.6 km atmospheric turbulence effects, an overall decrease in the distribution of coupling efficiency is observed compared to the case without atmospheric turbulence effects.
This decrease can be attributed to the inherent beam wander caused by atmospheric turbulence as the beam propagates through the atmosphere.
In the case of simulating 2.6 km atmospheric turbulence effects without stabilization, the distribution of coupling efficiency exhibits a mean of 0.405 and a standard deviation of 0.141.
However, with the application of stabilization technique, the mean value significantly improves to 0.616, accompanied by a standard deviation of 0.032.
This indicates a substantial enhancement in both the mean value, amounting to 52 $\%$, and the standard deviation, increasing over 4.4 times compared to the case without stabilization technique.
Hence, the proposed beam-wander-correction technique in this study demonstrates effective performance even under conditions of fast atmospheric turbulence.
Figure \ref{fig7}(a) presents the variation of coincidence counts over time when employing the auto-coupling algorithm and stabilization technique with a polarization entangled photon-pair source (see FSO setup in Fig. \ref{fig4}).
In the absence of stabilization (red-colored data), the mean value of coincidence counts is 5135 cps, accompanied by a corresponding standard deviation of 808 cps.
However, with successful stabilization, the mean value increases to 5834 cps, and the standard deviation decreases to 461 cps, indicating an improvement in the mean value by approximately 14 $\%$ and a doubling of the standard deviation.
To obtain accurate information about the beam position, a detection speed should be faster than the major frequency range ($\sim$ 300 Hz) caused by atmospheric turbulence.
In the measurements based on SPCMs, the accumulation of photon counts is set to 1 Hz.
This results in a somewhat lower performance in terms of position stabilization compared to the classical source case utilizing fast photo-diode.  
In the case of the quantum source, since it is not possible to directly measure the number of photons in front of the fiber collimator, coincidence counts of coupled signal photons in the receiver and directly measured idler photons from the source are demonstrated.

Finally, to verify the improvement in the quality of the entangled photon-pair source through stabilization, the fidelity of the polarization-entangled state is measured using quantum state tomography (QST) \cite{PRA01James}, as shown in Fig. \ref{fig7}(b).
Here, a comparison is made between the fidelity obtained directly from the polarization-entangled source, the fidelity of the transmitted and received polarization-entangled state under the condition of 2.6 km atmospheric turbulence effects, and the fidelity achieved when the stabilization method is applied.
The fidelity of the polarization-entangled source is measured to be 0.97$\pm$0.01.
However, after indoor transmission and reception of the entangled states, including the influence of 2.6 km atmospheric turbulence, the fidelity decreases to 0.83$\pm$0.07.
Since the fidelity measurements are based on coincidence photon counts, in unstable atmospheric conditions, without sufficient beam stabilization, coincidence counts significantly decrease and exhibit substantial fluctuations.
This can lead to significant deviations in QST measurement for each basis, resulting in a decrease in a fidelity.
By applying the proposed stabilization technique, the fidelity is restored to 0.95$\pm$0.02.
The slight decrease in fidelity compared to the source after stabilization can be attributed to external noise and the free-space transmission channel losses.

\section{Conclusion}
In this paper, we propose a highly-enhanced active beam-wander-correction technique, comprising an SMF auto-coupling algorithm and a decoupled stabilization method.
The proposed technique effectively searches for an optimal optical beam path, starting from a completely misaligned initial point, and compensates for beam wander caused by atmospheric turbulence by precisely controlling the angle and position of the beam using two sets of FSMs and PSDs.
To evaluate its performance, we conduct experiments using a 60-m indoor FSO setup capable of simulating atmospheric turbulence effects.
We measure the distribution of coupling efficiency and directly demonstrate the significant enhancement in coupling efficiency achieved through the proposed technique, even under the condition of 2.6 km atmospheric turbulence.
For the laser source, the coupling efficiency distribution exhibits a mean value increase of over 50 $\%$ and a standard deviation improvement of more than 4.4 times.
Similarly, for the entangled photon-pairs, the mean value increases by over 14 $\%$, and the standard deviation experiences an approximately twofold improvement.
Moreover, for the polarization-entangled state, we observe that the fidelity, which had decreased due to atmospheric turbulence effect, is effectively restored to a value close to the fidelity directly measured from the source when the stabilization method is applied.
The results obtained from out study signify the substantial potential of our proposed technique in designing fiber coupling optical systems, not only for free-space quantum communications but also for high-speed laser communications.
By mitigating the adverse impact of atmospheric turbulence on beam wander, our approach offers a promising solution for improving the efficiency and reliability of various optical communication systems, contributing to the advancement of both quantum and classical laser communication technologies.

\section*{Acknowledgments} 
This work was supported by Agency for Defense Development.

\section*{Disclosures} 
The authors declare no conflicts of interest.

\section*{Data availability} 
The datasets generated and/or analyzed during this study are not publicly available due to the security policy of the Ministry of National Defense of South Korea but are available from the corresponding author upon reasonable request.

\end{document}